\documentclass[12pt]{iopart}

\usepackage{iopams}\usepackage{graphicx}
\begin{document}

\title[A. Hutem and C. Sricheewin]{Ground-state energy eigenvalue calculation of the quantum mechanical well $V(x)=\frac{1}{2}kx^{2}\;+\;\lambda {x^{4}}$ \\via analytical transfer matrix method}

\author{Artit Hutem and Chanun Sricheewin}

\address{Condensed Matter Theory Research Unit, The Tah Poe Academia Institute (TPTP)\\
Department of Physics, Naresuan University, Phitsanulok 65000, Thailand }
\ead{newchanun@yahoo.com}
\begin{abstract}
The analytical transfer matrix technique is applied to the Schr\"{o}dinger equation of symmetric quartic-well potential problem in the form  $V(x)=\frac{1}{2}kx^{2}+\lambda{x^{4}}.$ This gives quantization condition from which we can calculate the ground-state energy eigenvalues numerically. We also compare the results with those obtained from numerical shooting method, perturbation theory, and WKB method.
\end{abstract}


 \noindent     \submitto{\JPA}


\maketitle
\section{Introduction}
Quantum Mechanical bound-state problems have long been of interest to physicists. There exist several means to study them,  e.g. WKB approximation \cite{Schiff1968}, time-independent perturbation theory \cite{Schiff1968}, numerical shooting method \cite{Giordano1997}, finite element method \cite{Ram-mohan2002,Ka-oey2004}.
 Here we analyze one-dimensional problem using the
 analytical transfer matrix method (ATMM) devised by \cite{Cao2001}. This method originated from planar optical wave-guides \cite{Cao1999}, and tunneling \cite{Asanithi2003}. The main principle is to divide the domain into many tiny segments. Each segment possesses a constant potential. The concept of a transfer matrix then arises when we connect the wave functions at the boundary of two different potential levels.

\section{Single-stepped potential}

For a background , we consider the stepped potential with a particular energy eigenvalue as shown in Fig.1.
It is well-known that the wave functions are plane waves. The original idea stems from the fact that the wave functions and their first derivatives can generally be written as linear combination of the form

\begin{figure}[t]
    \begin{center}
    \includegraphics[width=8.0cm,height=7.0cm,angle=0]{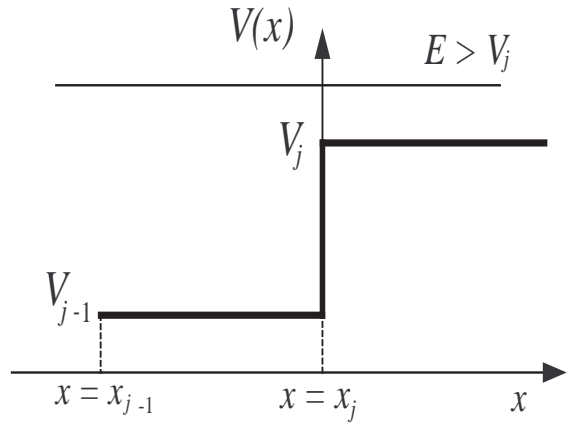}
    \end{center}
    \caption{The single-stepped potential}
    \end{figure}

\begin{eqnarray}
 \left[
   \begin{array}{c}
     \psi_{j-1}(x_{j-1}) \\
   \psi^{'}_{j-1}(x_{j-1})
   \end{array}
 \right]     =\left[
                \begin{array}{cc}
                  \alpha \;\; & \;\;\beta  \\
                 \gamma\;\; & \;\; \eta  \\
                \end{array}
              \right] \left[
   \begin{array}{c}
     \psi_{j-1}(x_{j}) \\
   \psi^{'}_{j-1}(x_{j})
   \end{array}
 \right],
 \end{eqnarray}
where $\alpha$, $\beta$, $\gamma$, $\eta$, are constants to be determined. We may substitute plane wave solution into
\begin{displaymath}
\psi_{j-1}(x_{j-1})=\alpha{\psi_{j-1}(x_{j})}\;+\;\beta{\psi^{'}_{j-1}(x_{j})},
\end{displaymath}
yielding
\begin{eqnarray}\nonumber
ae^{i\kappa_{j-1}(x_{j}-d)}+be^{-i\kappa_{j-1}(x_{j}-d)}&=&\alpha(ae^{i\kappa_{j-1}(x_{j})}+be^{-i\kappa_{j-1}(x_{j})})\\
&&+\beta(i\kappa_{j-1}ae^{i\kappa_{j-1}(x_{j})}-i\kappa_{j-1}be^{-i\kappa_{j-1}(x_{j})}),
 \end{eqnarray}
and into
\begin{displaymath}
\psi^{'}_{j-1}(x_{j-1})=\gamma{\psi_{j-1}(x_{j})}\;+\;\eta{\psi^{'}_{j-1}(x_{j})},
\end{displaymath}
giving
\begin{eqnarray}\nonumber
(i\kappa_{j-1}ae^{i\kappa_{j-1}(x_{j}-d)}-i\kappa_{j-1}be^{-i\kappa_{j-1}(x_{j}-d)})&=&\gamma(ae^{i\kappa_{j-1}(x_{j})}+be^{-i\kappa_{j-1}(x_{j})})\\
&&\!\!\!\!\!\!\!\!\!\!\!\!\!\!\!\!\!\!\!\!\!\!\!\!\!\!\!+\eta(i\kappa_{j-1}ae^{i\kappa_{j-1}(x_{j})}-i\kappa_{j-1}be^{-i\kappa_{j-1}(x_{j})}).
 \end{eqnarray}
 Here we regard $\mathbf{M}_{j}=\left[
                \begin{array}{cc}
                  \alpha\;\; &\;\; \beta \\
                 \gamma\;\; & \;\; \eta\\
                \end{array}
              \right] $
               as the matrix that transfers wave function and its first derivative from the position $x=x_{j-1}$ to $x=x_{j}$.
Solving Eq.(2) and Eq.(3) altogether, we obtain

\begin{eqnarray}\nonumber
\alpha&=&\cos(\kappa_{j-1}d),\;\;\;\beta=-\frac{1}{\kappa_{j-1}}\sin(\kappa_{j-1}d),\\
\gamma&=&\kappa_{j-1}\sin(\kappa_{j-1}d),\;\;\;\eta=\cos(\kappa_{j-1}d).
\end{eqnarray}
Substituting Eq.(4) into Eq.(1), we have
\begin{eqnarray}
 \left[
   \begin{array}{c}
     \psi_{j-1}(x_{j-1}) \\
   \psi^{'}_{j-1}(x_{j-1})
   \end{array}
 \right]     =\left[
                \begin{array}{cc}
                  \cos(\kappa_{j-1}d) & -\frac{1}{\kappa_{j-1}}\sin(\kappa_{j-1}d) \\
                 \kappa_{j-1}\sin(\kappa_{j-1}d) & \cos(\kappa_{j-1}d) \\
                \end{array}
              \right] \left[
   \begin{array}{c}
     \psi_{j-1}(x_{j}) \\
   \psi^{'}_{j-1}(x_{j})
   \end{array}
 \right],
 \end{eqnarray}
 where
 \begin{eqnarray}\nonumber
   \kappa_{j} &=& \frac{\sqrt{2m(E-V_{j})}}{\hbar}.\;\;j=1,2,...
 \end{eqnarray}
 On the other hand, in case the energy is lower than the stepped potential we would rather get
 \begin{eqnarray}
 \left[
   \begin{array}{c}
     \psi_{j-1}(x_{j-1}) \\
   \psi^{'}_{j-1}(x_{j-1})
   \end{array}
 \right]     =\left[
                \begin{array}{cc}
                  \cosh(\kappa_{j-1}d) & \frac{1}{\kappa_{j-1}}\sinh(\kappa_{j-1}d) \\
                 \kappa_{j-1}\sinh(\kappa_{j-1}d) & \cosh(\kappa_{j-1}d) \\
                \end{array}
              \right] \left[
   \begin{array}{c}
     \psi_{j-1}(x_{j}) \\
   \psi^{'}_{j-1}(x_{j})
   \end{array}
 \right].
 \end{eqnarray}

\section{The multi-stepped potential well}

\begin{figure}[t]
    \begin{center}
    \includegraphics[width=15.0cm,height=7.0cm,angle=0]{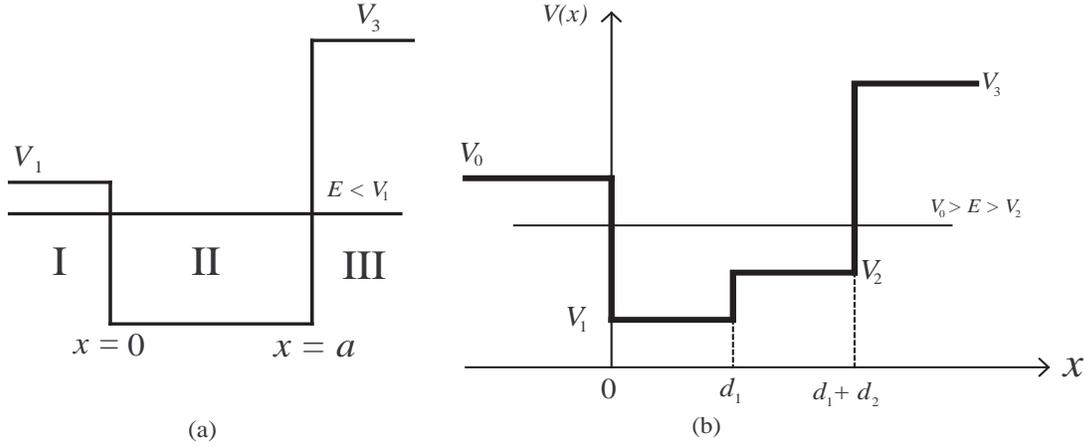}
    \end{center}
    \caption{(a) Asymmetric finite square-well potential in case of no phase contribution. (b) The stepped potential well with phase contribution.}
    \end{figure}

In order to solve the Schroedinger equation in each of three regions in Fig. 2(a), the boundary conditions due to the continuity of the wave function and its first derivative are applied at the boundaries of regions I, II, and III. In case the energy is less than both $V_{1}$ and $V_{3}$, the solution is given by \cite{Brennan1999}
\begin{equation}
k{a}=m\pi-\sin^{-1}[P_{2}]-\sin^{-1}[P_{1}],
\end{equation}
where
\begin{center}
 $P_{1}\equiv{\frac{\hbar{k}}{\sqrt{2mV_{1}}}}, \;\;\;\;P_{2}\equiv{\frac{\hbar{k}}{\sqrt{2mV_{3}}}},\;\;\;k\equiv\sqrt{\frac{2mE}{\hbar^{2}}}$.
 \end{center}
Consider the stepped potential well shown in Fig. 2 (b). We pay attention to the case of $E>V_{2}$. The wave function has the from of $A_{0}\exp{(P_{0}x)}$ in the regime $x<x_{0}$ and $A_{3}\exp{(-P_{3}x)}$ in the regime $x>d_{1}+d_{2}$. Between them, the wave function oscillates with different wavelengths whenever the particle moves from a constant potential to another step. We finally obtain the quantization rule as follows \cite{Cao2001}:
\begin{equation}
\kappa_{1}d_{1}+\kappa_{2}d_{2}+\Phi{(s)}=n\pi+\tan^{-1}\bigg[\frac{P_{0}}{\kappa_{1}}\bigg]+\tan^{-1}\bigg[\frac{P_{3}}{\kappa_{2}}\bigg],
\end{equation}
where\\
\begin{center}$\kappa_{j}\equiv{\frac{\sqrt{2m(E-V_{j})}}{\hbar}}; j=1,2,$\;\;\; $P_{j}\equiv{\frac{\sqrt{2m(V_{j}-E)}}{\hbar}}; j=0,3,$\\
$ \Phi(s)\equiv \Phi_{2}-\tan^{-1}\bigg[
\frac{\kappa_{2}}{\kappa_{1}}\tan(\Phi_{2})\bigg]$,\;\;\;$\phi_{2}=\acute{n}\pi+\tan^{-1}\bigg(\frac{P_{3}}{\kappa_{2}}\bigg)-\kappa_{2}d_{2},\;\;\acute{n}=0,1,2,...$\end{center}

\hspace{5mm} The second and third terms on the right-hand side of Eq.(8)are half-phase losses at the potential barriers $V_{0}$ and $V_{3}$, respectively. We observe that by setting  $V_{1}=V_{2}$, we obtain $\Phi(s)=0$, this phase contribution actually results from the interference of the scattered sub-waves between the potentials $V_{1}$ and $V_{2}$.

\section{An arbitrary potential-well function}

Since a continuous potential well may be viewed as a stack of thin films each of which possesses a constant potential, the above method
can therefore be extended to study an arbitrary one-dimensional potential well of the form shown in Fig. 3 below:

\begin{figure}[t]
    \begin{center}
    \includegraphics[width=9.0cm,height=7.0cm,angle=0]{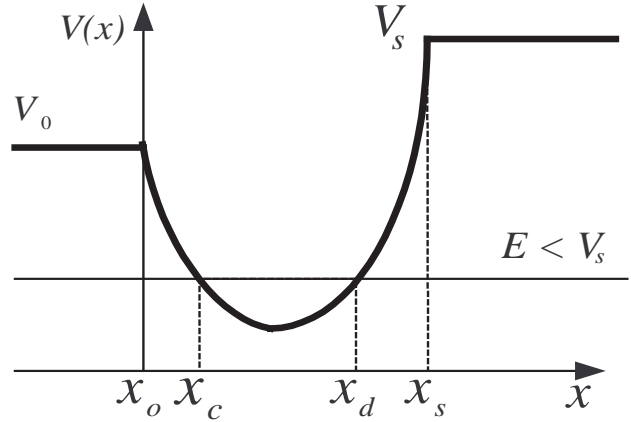}
    \end{center}
    \caption{An arbitrary potential well function $V(x)$}
    \end{figure}

\begin{figure}[t]
    \begin{center}
    \includegraphics[width=10.0cm,height=7.0cm,angle=0]{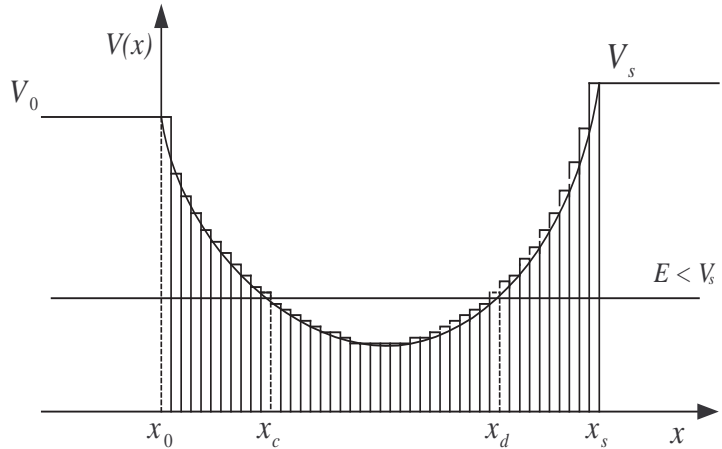}
    \end{center}
    \caption{A graph of an arbitrary potential well is equivalent to an assembling of tiny bars.}\label{D4}
    \end{figure}

We have also truncated the profile at $x=x_{0}$ and $x=x_{s}$. The transaction certainly affects the values of energy levels as compared to the
situation in idealized system. The effects will clearly be negligible if the potential at the transaction points is very much larger than
energies of relevant levels. Given that $x_{c}$ and $x_{d}$ are turning points, we divide the region $(x_{0},x_{c})$,$(x_{c},x_{d})$
and $(x_{d},x_{s})$ into $l,m$ and $n$ equal parts each of width $d$. According to the classically allowed or forbidden regions,
the multi-stepped potential corresponding to the $i$th, $j$th and $k$th section layers of the form are shown in Fig. \ref{D4}

The transfer matrices can be written as
\begin{equation}
\mathbf{M}_{i}=\left[
                 \begin{array}{cc}
                   \cosh(\alpha_{i}d) & -\frac{1}{\alpha_{i}}\sinh(\alpha_{i}d) \\
                   -\alpha_{i}\sinh(\alpha_{i}d) & \cosh(\alpha_{i}d) \\
                 \end{array}
               \right], i=1,2,...,l;
\end{equation}
\begin{equation}
\mathbf{M}_{j}=\left[
                 \begin{array}{cc}
                   \cos(\kappa_{j}d) & -\frac{1}{\kappa_{j}}\sin(\kappa_{j}d) \\
                    \kappa_{j}\sin(\kappa_{j}d)& \cos(\kappa_{j}d) \\
                 \end{array}
               \right], j=l+1,l+2,...,l+m;
\end{equation}
and
\begin{equation}
\mathbf{M}_{k}=\left[
                 \begin{array}{cc}
                   \cosh(\alpha_{k}d) & -\frac{1}{\alpha_{k}}\sinh(\alpha_{k}d) \\
                   -\alpha_{k}\sinh(\alpha_{k}d) & \cosh(\alpha_{k}d) \\
                 \end{array}
               \right], k=l+m+1,...,l+m+n;
\end{equation}
where
\begin{eqnarray}\nonumber
  \alpha_{i} &=& \frac{\sqrt{2m[V(x_{i})-E]}}{\hbar} \\
  \kappa_{j} &=& \frac{\sqrt{2m[E-V(x_{j})]}}{\hbar} \\ \nonumber
  \alpha_{k} &=& \frac{\sqrt{2m[V(x_{k})-E]}}{\hbar}.
\end{eqnarray}
Applying the boundary conditions at $(x=x_{0})$ and $(x=x_{s})$ yields
\begin{equation}
\left[
  \begin{array}{c}
    \psi(x_{0}) \\
    \psi'(x_{0}) \\
  \end{array}
\right]=\bigg[\prod_{i=1}^{l}\mathbf{M}_{i}\bigg]\bigg[\prod_{j=l+1}^{l+m}\mathbf{M}_{j}\bigg]\bigg[\prod_{k=l+m+1}^{l+m+n}\mathbf{M}_{k}\bigg]\left[
                                                                                                                                                 \begin{array}{c}
                                                                                                                                                   \psi(x_{s}) \\
                                                                                                                                                  \psi'(x_{s}) \\
                                                                                                                                                 \end{array}
                                                                                                                                               \right],
\end{equation}
where the prime denotes differentiation with respect to $x$. As known, the wave functions in the region $x<x_{0}$ is $A_{0}e^{P_{0}(x-x_{0})}$
and wave function in the region $x>x_{s}$ is $A_{s}e^{-P_{s}(x-x_{s})}$ where
\begin{eqnarray}\nonumber
  P_{0} &=& \frac{\sqrt{2m[V_{0}-E]}}{\hbar};x=x_{0},  \\
P_{s} &=& \frac{\sqrt{2m[V_{s}-E]}}{\hbar};x=x_{s}. \\ \nonumber
\end{eqnarray}
 $A_{0}, A_{s}$ are the amplitude coefficients to be determined. From Eq. (12) the solution for $j=l+m$ by using similar algebra manipulation as developed in the reference is given by \cite{Cao1999}
\begin{equation}
\kappa_{l+m}d=n_{l+m}\pi+\tan^{-1}\bigg[\frac{P_{l+m+1}}{\kappa_{l+m}}\bigg]-\Phi_{l+m}.
\end{equation}
Summing all indices $j$, we have
\begin{equation}
\sum_{j=l+1}^{l+m}\kappa_{j}d+\Phi(s)=\mathcal{N}\pi+\tan^{-1}\bigg[\frac{P_{l}}{\kappa_{l+1}}\bigg]+\tan^{-1}\bigg[\frac{P_{l+m+1}}{\kappa_{l+m}}\bigg], \;\;\mathcal{N}=0,1,\ldots
\end{equation}
where
\begin{equation}
\Phi(s)=\sum_{j=l+1}^{l+m-1}\bigg[\Phi_{j+1}-\tan^{-1}\bigg[\frac{\kappa_{j+1}}{\kappa_{j}}\tan\Phi_{j+1}\bigg]\bigg],
\end{equation}
and
\begin{eqnarray}\nonumber
  \Phi_{j} &=& \tan^{-1}\bigg(\frac{P_{j}}{\kappa_{j}}\bigg),\;\;
  P_{j} = \kappa_{j}\tan\bigg[\tan^{-1}\bigg(\frac{P_{j+1}}{\kappa_{j}}\bigg)-\kappa_{j}d\bigg].
\end{eqnarray}
\hspace{5mm}$\Phi(s)$ is the \textbf{phase contribution} devoted by the scattered sub-wave. $P_{l}$ and $P_{l+m+1}$ are equivalent exponential decaying coefficients corresponding to the regions $x<x_{c}$ and $x>x_{d}$, respectively, $P_{i}$ is momentum that transfers energy at a stack of thin films interval $i=1,2,\ldots,l$ and $P_{k}$ is momentum that transfers energy at a stack of thin films interval $k=l+m+1,l+m+2,\ldots,l+m+n$. We now investigate the half-phase losses at the turning point, i.e., the two terms $\tan^{-1}\bigg[\frac{P_{l}}{\kappa_{l}}\bigg]$ and $\tan^{-1}\bigg[\frac{P_{l+m+1}}{\kappa_{l+m}}\bigg]$ in Eq. (16). It is obviously clear that as $l,m,$ and $n\rightarrow{\infty}, d\rightarrow{0}$, we have $\kappa_{l+1}=0$ at $x_{l+1}$ and $\kappa_{l+m}=0$ at $x_{l+m}$. $P_{l}$ and $P_{l+m+1}$ are positive and finite for bound states so that the half-phase losses at the turning point have  the value $\frac{\pi}{2}$.
 We thus obtain a quantization condition as the width of the section layers $d$ tends to zero, i.e.
\begin{equation}
\int_{x_{c}}^{x_{d}}\kappa(x)dx+\Phi(s)=(\mathcal{N}+1)\pi,\;\;\mathcal{N}=0,1,2,\ldots .
\end{equation}

\section{Numerical Results}
We now pay attention to the particular problem of calculating ground-state energy of the potential-well
\begin{equation}
V(x)=\frac{1}{2}m\omega^{2}x^{2}+\lambda{x^{4}}
\end{equation}
where $\lambda$ is a positive constant.
After introducing new variable and parameter as $\beta\equiv\frac{2\lambda \hbar}{m^{2}\omega^{3}}$, $\xi=\alpha{x}$, $\alpha=\big(\frac{m\omega}{\hbar}\big)^{\frac{1}{2}}$, Schr$\ddot{o}$dinger equation is transformed to

\begin{equation}
\frac{d^{2}\psi(\xi)}{d\xi^{2}}+[\varepsilon-\xi^{2}-\beta\xi^{4}]\psi(\xi)=0.
\end{equation}
Also, the potential in term of new variable and parameter is given by
\begin{equation}
V(\xi)=\xi^{2}+\beta{\xi^{4}}.
\end{equation}
Due to this, the principal phase in the quantization condition Eq. (18) reduces, for the sake of computation, to

\begin{equation}
\int_{x_{c}}^{x_{d}}\kappa(x)dx=\int_{\xi_{c}}^{\xi_{d}}\sqrt{\varepsilon-\xi^{2}-\beta\xi^{4}}\;\;d\xi,
\end{equation}
where $\xi_{c}=\big(\frac{m\omega}{\hbar}\big)^{\frac{1}{2}}x_{c}$, $\xi_{d}=\big(\frac{m\omega}{\hbar}\big)^{\frac{1}{2}}x_{d}$.
The ATMM gives a quantization rule similar to that obtained from WKB method, the only difference is that for ATMM the phase contribution can be calculated with slight complication whereas WKB result may be equivalent to the ATMM case by setting the phase contribution equal to $\frac{\pi}{2}$.
We have calculated the ground-state energy eigenvalues under this potential as shown in Table 1.

\begin{table}
\caption{Comparison of ground-state energy obtained from ATMM., standard WKB.,
 $1^{st}$ order perturbation, and numerical shooting method.}
\begin{tabular*}{\textwidth}{@{}l*{15}{@{\extracolsep{0pt plus
12pt}}l}}
\br
$\beta\equiv\frac{2\hbar\lambda}{m^{2}\omega^{3}}$ & ATMM  &  WKB  & $1^{st}$P.T.$^{5}$ &NSM$^{5}$    \\
& (${\hbar{\omega}}/{2}$) & (${\hbar{\omega}}/{2}$) & (${\hbar{\omega}}/{2}$)&(${\hbar{\omega}}/{2}$) \\ 
\mr  
0.1\hphantom{00} & \hphantom{0}1.06640625 & \hphantom{0}1.03515625 & 1.07500000&1.06451896 \\
0.2\hphantom{00} & \hphantom{0}1.12131250 & \hphantom{0}1.06656250 & 1.15000000&1.11740450 \\
0.3\hphantom{00} & \hphantom{0}1.16906250 & \hphantom{0}1.09515625 & 1.22500000&1.16304919 \\
0.4\hphantom{00} & \hphantom{0}1.21234375 & \hphantom{0}1.12171875 & 1.30000000&1.20371079 \\
0.5\hphantom{00}  & \hphantom{0}1.25156250 & \hphantom{0}1.14687550 & 1.37500000& 1.24065919\\
0.6\hphantom{00} & \hphantom{0}1.28781250 & \hphantom{0}1.16934375 & 1.45000000&1.27469829 \\
0.7\hphantom{00} & \hphantom{0}1.32171875 & \hphantom{0}1.19106875 & 1.52500000&1.30637709 \\
0.8\hphantom{00} & \hphantom{0}1.35390625 & \hphantom{0}1.21203125 & 1.60000000&1.33609065 \\
0.9\hphantom{00} & \hphantom{0}1.38359375 & \hphantom{0}1.23187550 & 1.67500000&1.36413529 \\
1.0\hphantom{00} & \hphantom{0}1.41183595 & \hphantom{0}1.25156250 & 1.75000000 &1.39073967\\
\br
\end{tabular*}
\end{table}

\section{Conclusion}
 The NSM is generally regarded as one of the most efficient methods that give accurate results because it integrates the Schr\"{o}dinger equation directly, though in numerical sense. From table1, we see that the ATMM we have adopted from Cao et al \cite{Cao2001} gives outstandingly better results for ground state than those obtained from $1^{st}$ order perturbation theory and the typical WKB method for every value of $\lambda$ (or $\beta$) under this quartic single well potential. In our viewpoint, the complication of the ATMM computation lies at the evaluation of phase contribution, particularly when the potential is not of very simple form so that we cannot obtain analytical expression from the quantization rule but numerical one instead, in which we employ iteration method to work them out. In fact, the ATMM is claimed by Cao et al \cite{Cao2001} to give exact formalism of quantization rule without any approximation. One slight disagreement is emphasized here in that the truncation of the bound potential at $x_{0}$,$x_{s}$ in fact indicates a kind of approximation. Anyhow, this estimation seems to be a reasonable one that does not considerably affect the energy calculation especially when we choose $x_{s}$ far away enough from the RHS turning point.

\section*{Acknowledgements}

 We would like to thank  Burin Gumjudpai and Nattapong Yongram for their useful discussions. This work is supported by Naresuan Faculty of Science Research Scheme and the Condensed Matter Theory Research Unit of the Tah Poe Academia Institute,
Department of Physics, Naresuan University.

\end{document}